\documentclass[unknownkeysallowed]{pkas_ss}

\def\year{2014} 
\def\month{March} 
\def\volume{00} 
\def\issue{0} 
\def\beginpage{1} 
\def\endpage{99} 
\def\received{---} 
\def\accepted{---} 
\date{Received \received ; accepted \accepted}

\title{
NEP-AKARI: Evolution with redshift of dust attenuation in  8 $\mu$m selected galaxies
}

\author[1]{V.Buat}
\author[2]{N. Oi}
\author[1]{D. Burgarella}
\author[3,4]{K.Malek}
\author[2]{H. Matsuhara}
\author[2]{K. Murata}
\author[5]{S. Serjeant}
\author[6]{T.T. Takeuchi}
\author[7]{M.Malkan}
\author[6]{C. Pearson}
\author[2]{T. Wada}

\affil[1]{Aix-Marseille  Universit\'e,  CNRS, LAM (Laboratoire d'Astrophysique de Marseille) UMR7326,  13388, Marseille, France; \email{veronique.buat@lam.fr}}
\affil[2]{Institute of Space and Astronautical Science, Japan
Aerospace Exploration Agency, Sagamihara, Kanagawa 229-8510, Japan}
\affil[3]{Division of Particle and Astrophysical Science, Nagoya University, Furo-cho, Chikusa-ku, Nagoya 464-8602, Japan}
\affil[4]{National Centre for Nuclear Research, ul. Hoza 69, 00-681 Warszawa, Poland }
\affil[5]{ The Open University, Milton Keynes, MK7 6AA, UK}
\affil[6]{Graduate School of Science, Nagoya University, Nagoya, Aichi
464-8602, Japan }
\affil[7]{ University of California, Los Angeles, CA 90095-1547, USA}
\affil[8]{ Rutherford Appleton Laboratory Oxon, OX11 0QX, UK }

\def\corrauthor{
V. Buat
}

\def\runningauthor{
Buat et al.
}

\def\runningtitle{
Dust attenuation in  8 $\mu$m selected galaxies
}

\def\keywords{
Galaxies: IR--- Galaxies: evolution
}

\def\abstracttext{
We built a 8 $\mu$m selected sample of galaxies in the NEP-AKARI field by defining 4
redshift bins with the four AKARI bands at 11, 15, 18 and 24 microns ($0.15<z<0.49$,
$0.75<z<1.34$, $1.34<z<1.7$ and  $1.7<z<2.05$) . Our sample contains 4079 sources, 599 are securely detected with
$Herschel$/PACS. Also adding ultraviolet (UV) data from $GALEX$, we fit the spectral energy distributions using the physically motivated code
CIGALE to extract  the  star formation rate, stellar mass, dust
attenuation and the AGN contribution to the total infrared  luminosity ($L_{\rm IR}$). We discuss  the impact of the adopted
attenuation curve and  that of the wavelength coverage  to estimate these physical parameters. We  focus on  galaxies with a luminosity close the characteristic $L_{\rm IR}^*$ in the different redshift bins to study
the evolution with redshift of the dust attenuation in these galaxies.
}

\begin{document}
\pkashead 


\section{Introduction}
The strong impact of dust attenuation and the need to estimate it as accurately as possible  has led to numerous investigations on   samples of star forming galaxies. Quite naturally these  studies focused on  galaxies selected either in their UV rest-frame or from their emission in recombination lines and whose detection is strongly affected by the attenuation by dust. The aim of the present work is to perform a complementary study by starting with an  infrared (IR)  selection since dust re-emits  the   absorbed energy  of young stars   in this wavelength range. Most of the star formation in the universe can be securely measured in  IR up to redshift $\sim 2$ since dust emission dominates the stellar emission. The question we want to answer is: what is the amount of dust attenuation in these galaxies and how does it compare to the global evolution of dust attenuation in the universe and to the trends found in optically selected samples?\\
\section{Sample selection}
The infrared space telescope $AKARI$ carried out a deep survey of the North Ecliptic Pole (hereafter NEP-deep) with all the filters of the InfraRedCamera (IRC). We take advantage of the continuous filter coverage in the mid-IR   to build a 8$\mu$m rest-frame selection following the strategy of \citet{goto10}. Using the S11, L15, L18W and L24 filters we select galaxies from $z=0.15$ to $z=2.05$. The NEP-$AKARI$ field was also observed by $Herschel$ with the PACS and SPIRE instrument (P.I. S. Serjeant)  and  by $GALEX$ (P.I. M. Malkan). Given the small number of sources detected with SPIRE we only consider PACS data. UV data are also  added to galaxies with redshift lower than 0.925 in order that the NUV filter at 230 nm corresponds to rest-frame wavelength larger than 120 nm).  Fluxes  at 100 and 160 microns  from the PACS images  and at 230 nm in the GALEX images were  measured with DAOPHOT (Mazyed et al., in preparation). Combining the near to mid IR catalogue of \citet{murata13} with the optical data of \citet{oi14} and with $GALEX$ and $Herschel$/PACS detections, we are able to build the UV to IR spectral energy distributions (SED) of our selected sources. The photometric redshifts are  from \cite{oi14}. The SEDs are analysed with the SED fitting code CIGALE in order to measure physical parameters such as dust attenuation, SFR or stellar masses.  The amount of dust attenuation and its evolution in redshift is then measured for our selected sources and compared to the results found in UV selections.\\
\section{Fitting the spectral energy distributions}
\begin{table}[t!]
\caption{Values of input parameters used for the SED fitting with CIGALE, see text for details}
\label{tab:parameters}
\centering
\begin{tabular}{l  l}
\hline
Parameter & Range \\
\hline\hline
 $E(B-V)$  &0.02-1\\
Attenuation law & B12,C00, SMC\\
IR templates, $\alpha$& 1-3\\
AGN fraction,$\rm frac_{AGN}$& 0-0.5\\
\hline
{Stellar populations} \\
\hline
age (old stellar population)  $t_f$ &  2-11\,Gyr \\
$e$-folding rate  $\tau$ &  1-5\,Gyr \\
age  (young stellar population) $t_{\rm ySP}$ &  50-500\,Myr \\
stellar mass fraction   $f_{\rm ySP}$ &  0.01-0.2\\
\hline
\end{tabular}
\end{table}

The spectral energy distributions (SEDs) are fitted  with the new version of the CIGALE code (Code Investigating GALaxy Emission)\footnote{http://cigale.lam.fr} developed with PYTHON.  CIGALE combines a UV-optical stellar SED with a dust component emitting in the IR and fully conserves the energy balance between the dust absorbed stellar emission and its re-emission in the IR. We refer to Burgarella et al. (in preparation) and Boquien et al. (in preparation) for a detailed description of the new version code.  The earlier version of CIGALE is described in \cite{noll09}. Here we only describe  the assumptions and choices specific to the current study.  The choice of the models and parameters is the result of a long process of optimization which will be fully discussed in Buat et al. (in preparation).\\
For the star formation we adopt the fiducial  model of  \citet{buat14} which consists of two stellar populations:  a recent stellar population with a constant SFR on top of an older stellar population created with an exponentially declining SFR ($e$-folding rate  $\tau$). Ages of the older stellar population and young component ($t_f$  and  $t_{\rm ySP}$ respectively)  are free parameters. The two stellar components are linked by their stellar mass fraction $f_{\rm ySP}$. \\
 We can run  CIGALE with different scenarios for the dust attenuation. 
Our fiducial attenuation recipe  is the one we obtained by studying high redshift galaxies  \cite{buat12}, hereafter B12), it  is close to the LMC2 extinction curve with a UV bump of moderate amplitude and a rather steep increase in the UV. We also consider the   grayer attenuation law of  \cite{calzetti00} (hereafter C00) and  an SMC extinction curve (hereafter SMC). The amount of dust extinction is measured with $E(B-V)$. The main parameters and the range of the input values are reported in Table 1.\\
The templates used for the dust re-emission in the IR   come from \cite{dale14} (used without any AGN contribution). With CIGALE,  we can add the contribution of an AGN to the SED. The adopted templates are those of the  \cite{fritz06} library. Our fiducial fits are performed with two models of type 2 AGN with a low or a high optical depth at 9.7 $\mu$m.\\
The reduced $\chi_r^2$ distribution is  found  very  good: for $93\%$ of the sources the minimum value of $\chi_r^2$ is lower than 5, with a similar result for each redshift bin and for galaxies detected with PACS  or not. The AGN fraction is of the order of $10\%$ in average for the whole sample. \\

\begin{table}[t!]
\caption {Impact of the choice of the attenuation law and of considering IR data on  the SFR and stellar masses ($M_{\rm star}$) determinations. The first part of the table summarises  the comparison between the 3 different attenuation laws considered in this work, in the second part of the table  the  results of the fits with and without IR data are compared  for each of the 3 attenuation laws}
\begin{tabular}{ccc} 
\hline
Models   & $\log(M_{\rm star})$& $\log(SFR)$\\
                &dex&dex\\
\hline
with IR &&\\
\hline
  C00-B12 & 0.09$\pm 0.08$&-0.03$\pm 0.14$ \\
  SMC-B12& -0.02 $\pm 0.05$ &-0.04  $\pm 0.10$\\
  C00-SMC&0.11$\pm 0.10$&0.01 $\pm 0.19$\\
  \hline
  without  IR - with IR &&\\
  \hline
  C00-C00& -0.03 $\pm 0.11$ & 0.14 $\pm 0.30$ \\
  B12-B12& -0.06 $\pm 0.14$ & 0.17  $\pm 0.36$\\
  SMC-SMC & -0.06 $\pm 0.14$& 0.11 $\pm 0.38$ \\
\hline\end{tabular}
\end{table}

We explore the influence of the adopted attenuation curve and of the presence or not of IR ($\rm \ge 8 \mu m$) data on the determination of  the  SFR and stellar mass determination. The  results are summarized in Table 2. In IR selected galaxies, the SFR is essentially constrained by the infrared emission from dust and does not depend much on the attenuation law. The stellar masses are found larger for the C00 attenuation law. The C00 curve exhibits the steepest increase in the UV  and thus  gives the lowest attenuation in the UV and the highest one in the V band since the total attenuation is constrained by $L_{\rm IR}$. Nevertheless the difference does not exceed $\sim 0.1$ dex.\\

 Omitting or not  IR data  has large consequences on  the measure of  the SFRs  whereas the stellar masses remain unaffected. This variation is due to the uncertainty on dust attenuation when IR data are missing.   The average  difference can reach 50$\%$  with a large dispersion (larger than a factor 2). 
 
 \section{Evolution of dust attenuation}

\begin{figure}
   \centering
  \includegraphics[width=\columnwidth]{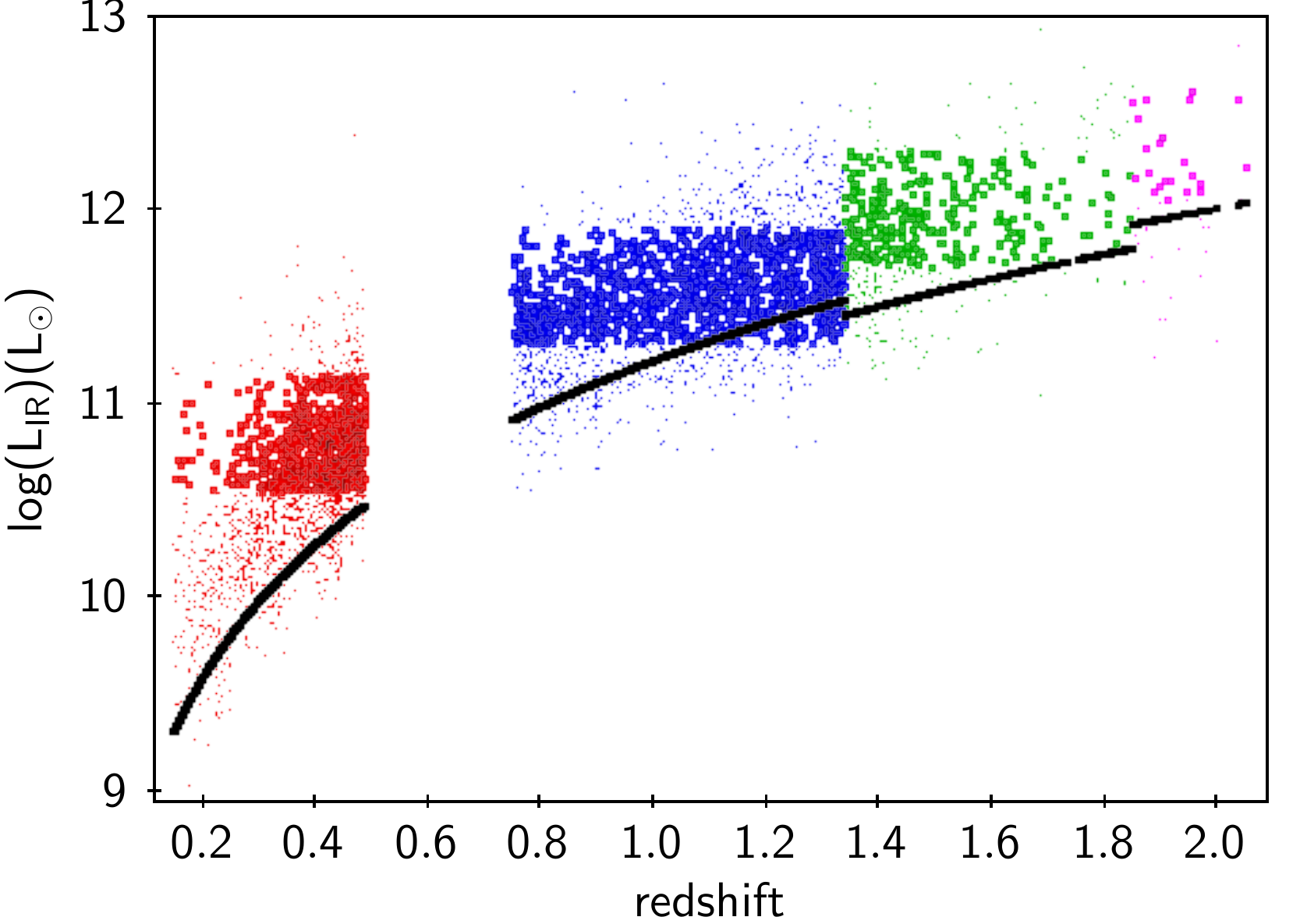}
   \caption{ Redshift and luminosity distributions of the sources. The total IR luminosity $L_{\rm IR}$ is an output of the SED fitting.  The small dots  represent the whole sample and the filled circles the galaxies selected around $L_{\rm IR}^*$ within each redsfhift bin. The adopted values of $\log(L_{\rm IR}^*)$ for the 4 redshift bins are  10.84, 11.6, 12 and 12.35 $L_{\odot}$, the bin size is 0.6 dex. Detection limits at 5$\sigma$  are reported as a black line.}
              \label{Ldust-z}%
    \end{figure}

Our aim it to measure dust attenuation in galaxies selected in a similar way at different redshifts.
The combination of the different IRC filters led us  to perform a  $8\mu$m rest-frame selection. In  Fig.\ref{Ldust-z}, $L_{\rm IR}$  (obtained with CIGALE)  is reported against  the photometric redshift of the sources. We have estimated the detection limit at 5$\sigma$ using the flux limits of \citet{murata13} for each IRC filter used to define the sample  and the average ratio between $L_{\rm IR}$ (found by SED fitting) and the monochromatic luminosity  in the IRC filter corresponding to the redshift selection.  It is clearly seen that we cannot follow  galaxies with a similar   $L_{\rm IR}$ over the full redshift range since the limit in luminosity increases sharply with z. We decide to study galaxies sampling the same domain of the luminosity function at the different redshifts. We    select the galaxies dominating the luminosity function and therefore the luminosity density at a given redshift. At this aim, we take the total IR luminosity functions of \citet{magnelli13}. In each redshift bin, we  select galaxies with an IR luminosity inside a bin of 0.6 dex centered on  the characteristic IR luminosity $L_{\rm IR}^*$ corresponding to  the transition luminosity of the double power law function.  The selection is also represented in Fig.\ref{Ldust-z}. It can be checked that the selected sources lie above the 5$\sigma$ detection limit (except a small overlap at the end of the second redshift bin), the number of selected sources in each bin is 729, 1145, 284 and 24 from bin 1 to bin 4. 
The variation with redshift  of the  attenuation is plotted in Fig.2. \\

\begin{figure}
  \includegraphics[width=\columnwidth]{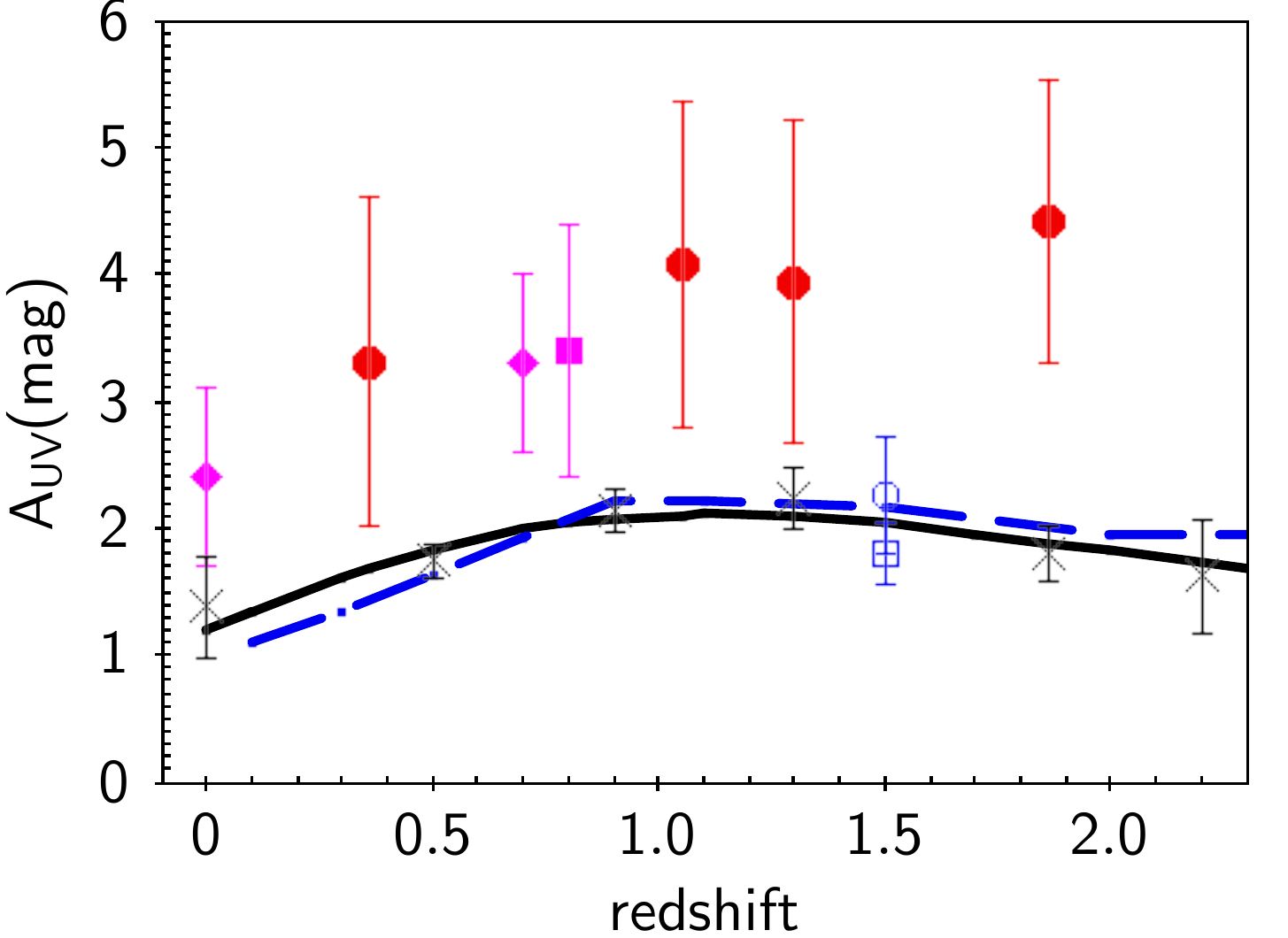}
 \caption{Dust attenuation in UV ($A_{\rm UV}$) plotted against   redshift . Red filled circles: present study with the   selection around $L_{\rm IR}^*$ in each redshift bin.   Magenta  filled lozenges:  \citet{buat06,buat08};   magenta filled  square: \cite{choi06}. The blue symbols  refer to   UV and $\rm H_{\alpha}$ line selections. Blue dashed line:  \cite{cucciati12}; blue empty square: \citet{heinis13}; blue empty circle: \citet{ibar13} . The global estimates of   \cite{burgarella13} are plotted with black crosses and their average fit with  a solid line  The  dispersion of each measurement is reported with a vertical bar.}
              \label{Afuv-z}%
    \end{figure}

We can first  compare our measurements to other values already obtained for IR selected galaxies. At $z=0$ \citet{buat06} combined $GALEX$ and $IRAS$ data and derived volume averaged measures of the attenuation. For   a luminosity  $L_{\rm IR}^*\sim 10^{10.5} L_{\odot}$ \citep{sanders03} the average UV attenuation  $A_{\rm UV} = 2.4\pm 0.7$ mag. \citet{buat08} measured the attenuation of Luminous IR Galaxies ($L_{\rm IR} > 10^{11} {\rm L_{odot}}$) at $z=0.7$ by combining 
{\it Spitzer} and $GALEX$ data and found a mean attenuation of 3.33 mag with a dispersion similar to that found at $z=0$.  \cite{choi06} also measured dust attenuation  in galaxies of the {\it Spitzer} First Look Survey selected in mid-IR at $z=0.8$ by comparing SFR measured with the strength of emission lines and $L_{\rm IR}$.  We apply their relation between $A_{\rm V}$ and $L_{\rm IR}$ to the average value of $L_{\rm IR}$ of our sample at the same redshift    and get $A_{\rm V}=2.33$ mag. The visual extinction  can then be  translated to an attenuation in the UV continuum   as explained in \cite{buat08} giving $A_{\rm UV} = 3.4$ mag. The dispersion is directly measured on Fig.12  of \cite{choi06}. All these previous measurements  are over-plotted on  Fig.\ref{Afuv-z} and are found consistent with the ones   obtained in the present work which extents the analysis at higher redshift.  The attenuation in UV for $L_{\rm IR}^*$ galaxies  is found to increase  with redshift from 2.4 mag at $z=0$ to 4.4 mag at $z\sim 2$.\\
We now compare the redshift evolution of the attenuation with measurements obtained in samples not   selected  in IR. \citet{burgarella13} measured the global attenuation in the universe by comparing the IR and UV luminosity densities. Their result is plotted in Fig.\ref{Afuv-z}. Measures of dust attenuation are also performed in UV selected samples. \cite{cucciati12}   derived dust attenuation for each galaxy of their UV selection through SED fitting (without IR data), the variation they found is in close agreement with the results of \citet{burgarella13}.  \cite{heinis13} measured the average attenuation of a UV selection at $z\sim 1.5$ in the COSMOS field by stacking $Herschel$/SPIRE images. The average value found for their whole selection is consistent with the one derived by \cite{cucciati12} at the same redshift ($z=1.5$). \citet{ibar13} analyzed the IR properties of galaxies detected in their H$\alpha$ line (HIZELS project) at $z=1.46$. From a stacking analysis of {\it Spitzer}, {\it Herschel} and AzTEC images  they derived the average total IR luminosity of their sample and deduced  a  median attenuation  of $A_{\rm H\alpha}=1.2 \pm 0.2$ mag for their sample. This attenuation can be translated in UV using the recipe of \cite{calzetti97}. 
The corresponding value (2.23$\pm$0.4 mag)  is plotted in Fig.2  and is fully consistent with the other measures  performed in a UV selection or globally with the luminosity functions.\\

The amount of attenuation found for our IR selection is much higher than the values found either in a UV selection or globally in the universe. When performing our IR selection we select galaxies dominating the star formation at the redshifts we consider. These galaxies are found to experience a dust attenuation much higher than the average one at work in the universe or found in UV selected samples: the galaxies dominating the  star formation are not representative of the average attenuation  and metal production in the universe.  \\
Dust attenuation is found to  depend on the stellar mass  and the relation between these quantities is not found to significantly evolve  with redshift  \citep{ibar13,kashino13,heinis14}. We can  use these relations to account for the stellar mass distributions when we compare dust attenuation from different selections. Let us consider  the measure of \citet{heinis13} at  $z\sim1.5$:  the average stellar mass for our IR selection in the redshift bin 3 is $10^{10.9} M_{\odot}$. Using the relation found by \cite{heinis14}   between dust attenuation and $M_{star}$,  we find that the average attenuation increases from 1.8 to 3.5 mag, leading to  a value now consistent with the measures obtained in our IR samples. \\
The same exercise can be performed for the H$\alpha$ selected sample.  Using the relation of \cite{kashino13} between attenuation and stellar mass  for H$\alpha$ emitters we find $A_{\rm H\alpha}=1.96$ mag which translates to 3.64 mag in UV again  consistent with what is found for our IR selected samples of similar mass.\\
Hence the  stellar mass appear as the main driver to reconcile measures of dust attenuation in the different studies and explain the discrepancies between UV and IR selections as predicted by the empirical models of \cite{heinis14} and \cite{bernhard14}.
\acknowledgments

V. Buat and D. Burgarella acknowledge the support of offering visiting professor positions in 2013 from the Institute of Space and Astronautical Science. V. Buat is also supported by the grant of the Institut Universitaire de France.


\end{document}